\documentclass[11pt,
twocolumn,
 aip,
 jmp,%
 showkeys,
 amsmath,amssymb,
reprint,%
]{revtex4-1}

\usepackage[breaklinks=true,colorlinks=true,linkcolor=blue,urlcolor=blue,citecolor=blue]{hyperref}
\usepackage{graphicx}%
\usepackage{graphics}%
\usepackage{epstopdf}
\usepackage{dcolumn}%
\usepackage{soul,color}
\usepackage{bm}
\usepackage[mathlines]{lineno}
\expandafter\ifx\csname package@font\endcsname\relax\else
 \expandafter\expandafter
 \expandafter\usepackage
 \expandafter\expandafter
 \expandafter{\csname package@font\endcsname}%
\fi
\begin{document}
\abovedisplayskip=3pt
\belowdisplayskip=3pt
\abovedisplayshortskip=2pt
\belowdisplayshortskip=2pt


\title{\Large{Engineered magnetization and exchange stiffness in direct-write Co-Fe nanoelements}}
\author{S. A.~Bunyaev}
    \affiliation{Institute of Physics for Advanced Materials, Nanotechnology and Photonics (IFIMUP)/Departamento de F\'isica e Astronomia, Universidade do Porto, 4169-007 Porto, Portugal}
\author{B. Budinska}
    \affiliation{Faculty of Physics, University of Vienna, 1090 Vienna, Austria}
\author{R. Sachser}
    \affiliation{Physikalisches Institut, Goethe University, 60438 Frankfurt am Main, Germany}
\author{Q. Wang}
    \affiliation{Faculty of Physics, University of Vienna, 1090 Vienna, Austria}
\author{K. Levchenko}
    \affiliation{Faculty of Physics, University of Vienna, 1090 Vienna, Austria}
\author{S. Knauer}
    \affiliation{Faculty of Physics, University of Vienna, 1090 Vienna, Austria}
\author{A. V.~Bondarenko}
    \affiliation{Institute of Physics for Advanced Materials, Nanotechnology and Photonics (IFIMUP)/Departamento de F\'isica e Astronomia, Universidade do Porto, 4169-007 Porto, Portugal}
\author{M.~Urbanek}
    \affiliation{CEITEC BUT, Brno University of Technology, 61200 Brno, Czech Republic}
\author{K.~Y.~Guslienko}
    \affiliation{Division de Fisica de Materiales,
      Depto. Polimeros y Materiales Avanzados: Fisica, Quimica y Tecnologia,
      Universidad del Pais Vasco, UPV/EHU,
      20018 San Sebastian, Spain}
    \affiliation{IKERBASQUE, the Basque Foundation for Science,
     48009 Bilbao, Spain}
\author{A.~V. Chumak}
    \affiliation{Faculty of Physics, University of Vienna, 1090 Vienna, Austria}
\author{M. Huth}
    \affiliation{Physikalisches Institut, Goethe University, 60438 Frankfurt am Main, Germany}
\author{G.~N. Kakazei}
    \affiliation{Institute of Physics for Advanced Materials, Nanotechnology and Photonics (IFIMUP)/Departamento de F\'isica e Astronomia, Universidade do Porto, 4169-007 Porto, Portugal}
\author{O.~V. Dobrovolskiy}
    \affiliation{Faculty of Physics, University of Vienna, 1090 Vienna, Austria}
    \email{oleksandr.dobrovolskiy@univie.ac.at}
\date{\today}

\begin{abstract}
Media with engineered magnetization are essential building blocks in superconductivity, magnetism and magnon spintronics. However, the established thin-film and lithographic techniques insufficiently suit the realization of planar components with on-demand-tailored magnetization in the lateral dimension. Here, we demonstrate the engineering of the magnetic properties of CoFe-based nanodisks fabricated by the mask-less technique of focused electron beam induced deposition (FEBID). The material composition in the nanodisks is tuned \emph{in-situ} via the e-beam waiting time in the FEBID process and their post-growth irradiation with Ga ions. The magnetization $M_s$ and exchange stiffness $A$ of the disks are deduced from perpendicular ferromagnetic resonance measurements. The achieved $M_s$ variation in the broad range from $720$\,emu/cm$^3$ to $1430$\,emu/cm$^3$ continuously bridges the gap between the $M_s$ values of such widely used magnonic materials as permalloy and CoFeB. The presented approach paves a way towards nanoscale 2D and 3D systems with controllable and space-varied magnetic properties.
\end{abstract}


\maketitle

Magnonics -- the study of spin waves and their use in information processing systems -- has emerged as one of the most rapidly developing research fields of modern magnetism \cite{Kru10jpd,Dem13boo,Chu15nph,Wag16nan,Dob19nph,Tsy19boo,Gub19boo,Mah20jap,Wan20nel,Liy20jap}. The key challenges in modern magnonics are guiding and control of spin waves in 1D (e.g., magnonic crystals  \cite{Kra14pcm,Kak14apl,Chu17jpd,Zak20pcm}), 2D (e.g., magnonic circuits \cite{Mah20jap,Wan20nel}), and emerging 3D systems \cite{Kra08prb,Yan11apl,Gub19boo}. For steering of spin waves, one should change an external parameter such as magnetic field \cite{Chu09jpd,Liu13apl,Gol18afm,Dob19nph}, temperature \cite{Pol12apl,Dzy16apl,Vog18nsr} or alter the conduit shape \cite{Kra14pcm,Gru18prb,Dob19ami} and magnetization \cite{Bau18apl,Dav15prb,Vog18nsr,Whi19prb,Bal14nal,Urb18apl}. Variations of temperature and field have drawbacks of being energy-consuming, and their localized application is challenging to implement. A shape variation is free of these complications but it usually involves multiple nanofabrication steps and allows one to only coarsely alter the magnetization between the given material's value and zero in regions where no material is present. Among these approaches, magnetization variation has an advantage if being passive (no current or heat involved) and it can be strongly localized or gradient-tailored on purpose. Thus, \emph{in-situ} approaches for tuning magnetization in a broad range are strongly demanded.

\begin{figure*}[t!]
    \centering
    \includegraphics[width=0.55\linewidth]{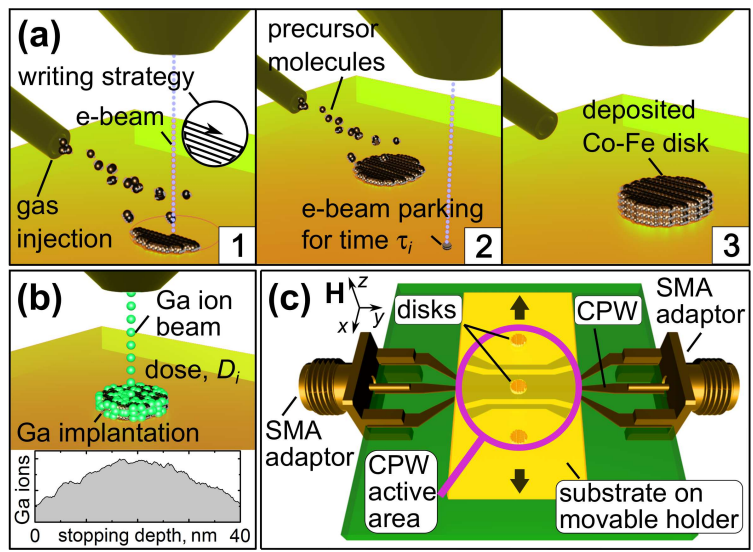}
    \caption{(a) Illustration of the FEBID process for the first series of disks: After each pass over the sample surface (1), the beam is parked outside of the disk for the given time $\tau_i$ (2). The writing process is continued until the desired disk thickness is achieved (3).
    (b) In the second series of measurements, a Co-Fe disk is irradiated by $30$\,keV Ga ions with different doses $D_i$.
    Inset: Simulated distribution of stopped Ga ions across the disk thickness.
    (c) Experimental geometry (not to scale). A substrate with a series of Co-Fe nanodisks is placed face-to-face to a gold coplanar waveguide (CPW) for spin-wave excitation in the out-of-plane bias magnetic field $\mathbf{H}$.
    }
    \label{f1}
\end{figure*}

In this regard, focused electron beam induced deposition (FEBID) can offer unique features which go beyond the state-of-the-art fabrication technologies of magnonics \cite{Hut18mee}. Firstly, the down to $10$\,nm lateral resolution (for selected materials, such as Co-Fe alloys \cite{Por15nan} used here) makes FEBID suitable for the fabrication of nanostructures with feature sizes comparable to modern complementary metal-oxide semiconducting (CMOS) technology. Secondly, the composition and magnetic properties of FEBID nanostructures can be tuned \emph{in-situ} by altering the writing strategy and via post-growth irradiation of structures with ions \cite{Lar14apl,Dob19ami} and electrons \cite{Dob15bjn}. In addition, FEBID is capable to fabricate complex-shaped 3D nano-architectures \cite{Fer20mat,Por19acs}, that make it the technique of choice for studies in superconductivity \cite{Dob11pcs,Dob18nac,Dob19pra}, magnetism \cite{Pac17nac,Str16jpd,She20cph} and magnonics \cite{Kra08prb,Yan11apl,Gub19boo}.

In a previous study, we observed the decrease of the magnetization $M_s$ and the exchange stiffness $A$ with reduction of the diameter of individual Co-Fe nanodisks \cite{Dob20nan}. The effect was attributed to the writing of smaller disks in a depleted-precursor regime which results in a lower metal content. Here, we use this effect to demonstrate on-demand engineering of the magnetization and exchange stiffness in individual Co-Fe nanodisks with a fixed radius $R = 500$\,nm and thickness $40$\,nm. Specifically, one series of nanodisks was fabricated using different e-beam waiting times in the FEBID process and another series of nanodisks was irradiated with different doses of Ga ions. The magnetization $M_s$ and exchange stiffness $A$ of the disks were deduced from ferromagnetic resonance (FMR) measurements, employing a recently developed spatially-resolved approach \cite{Dob20nan}. We demonstrate that with an increase of the e-beam waiting time, $M_s$ of the disks reaches $1430$\,emu/cm$^3$, which is by a factor of two larger than $M_s$ of the disks irradiated with Ga ions. Thus, the combination of these two approaches provides access to the fabrication of geometrically uniform magnonic conduits with a drastic variation of saturation magnetization.

The circular Co-Fe disks were fabricated by FEBID in a high-resolution dual-beam scanning electron microscope (SEM: FEI Nova NanoLab 600) employing HCo$_3$Fe(CO)$_{12}$ as precursor gas \cite{Por15nan,Kum18jpc}. 
Two series of samples were fabricated. The first series of samples is a set of four disks deposited onto a Si/SiO$_2$\,(200\,nm) substrate, written with different beam waiting times. Namely, after each pass of the electron beam over the disk surface, the beam was ``parked'' for the time $\tau$ varied from $\tau_0=0$ to $\tau_3=50$\,ms outside of the disk. The essential steps of the writing process are illustrated in Fig. \ref{f1}(a). The substrate with the disks was mounted onto a translational stage for their face-to-face positioning under the $2\,\mu$m-wide and $6\,\mu$m-long active part of an Au coplanar waveguide (CPW), Fig. \ref{f1}(c). The CPW was prepared by e-beam lithography from a $55$\,nm-thick Au film dc-magnetron-sputtered onto a Si/SiO$_2$\,(200\,nm) substrate with a $5$\,nm-thick Cr buffer layer. The CPW was covered with a $5$\,nm-thick TiO$_2$ layer for electrical insulation from the disks.

The second series of samples refers to four measurements of a disk written with $\tau_0=0$ on the CPW and irradiated with $30$\,keV Ga ions up to a cumulative dose $D$ of $15$\,pC/$\mu$m$^2$ in steps of $5$\,pC/$\mu$m$^2$, Fig. \ref{f1}(b). SRIM simulations of the distribution of $30$\,keV Ga ions implanted in the Co-Fe disks indicate that it has a gentle-dome shape spreading through the entire disk thickness, with the largest number of stopped Ga ions in the depth range from $13$\,nm to $28$\,nm, see the inset in Fig. \ref{f1}(b). Ferromagnetic resonance measurements on both sample series were taken at the fixed frequency $9.85$\,GHz with magnetic field oriented perpendicularly to the disk plane, Fig. \ref{f1}(c).

For the analytical description of the field values of resonance peaks, we considered azimuthally symmetric spin-wave modes in a thin cylindrical ferromagnetic disk saturated in the out-of-plane direction by the biasing magnetic field $H$. In this case, the excited spin-wave eigenmodes can be described by Bessel functions of the zeroth order because of the axial symmetry of the samples. Details of the analytical theory can be found elsewhere \cite{Kak04apl}. This approach allows for the deduction of $M_s$ and $A$ with high precision.

\begin{figure}[t!]
    \centering
    \includegraphics[width=0.98\linewidth]{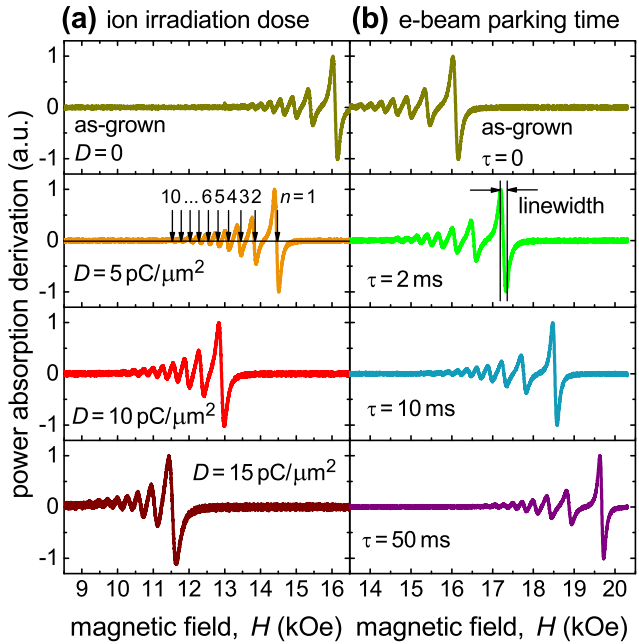}
    \caption{Experimentally measured spin-wave resonance spectra at $9.85$\,GHz for a series of $40$\,nm-thick Co-Fe disks with radius $R = 500$\,nm irradiated with Ga ions at different doses, as indicated (a) and deposited with different electron beam parking times (b). The resonance mode number $n$ and the peak-to-peak resonance linewidth are indicated.}
    \label{f2}
\end{figure}

\begin{figure}[t!]
    \centering
    \includegraphics[width=0.9\linewidth]{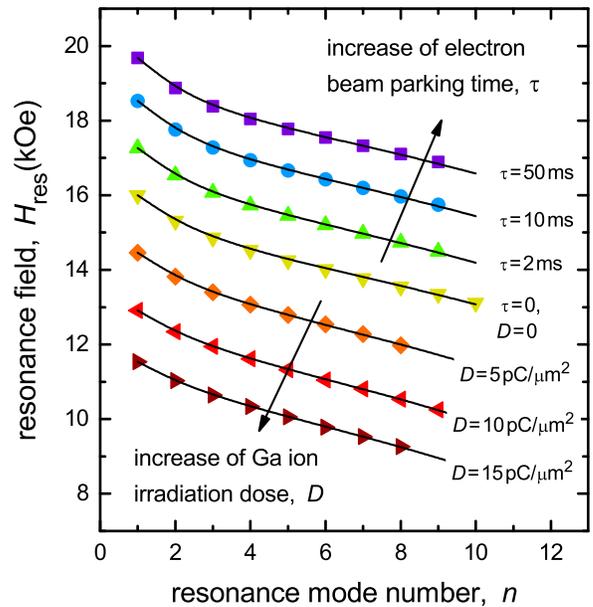}
    \caption{Dependences of the resonance field $H_\mathrm{res}$ on the spin-wave mode number $n$ for the disks irradiated with Ga ions at different doses and disks deposited with different parking times of the electron beam after each pass. Symbols: experiment. Solid lines: fits to the analytical theory \cite{Kak04apl} with the magnetization $M_\mathrm{s}$ and the exchange constant $A$ varied as fitting parameters, as reported in Fig. \ref{f4}, and the gyromagnetic ratio $\gamma/2\pi = 3.05$\,MHz/Oe.}
    \label{f3}
\end{figure}

\enlargethispage{1\baselineskip}
Figure \ref{f2} presents the experimentally measured spin-wave resonance spectra as a function of the out-of-plane magnetic field $H$ for the disks irradiated with different doses of Ga ions and the disks deposited with different electron beam waiting times. In all cases the most intense resonance peak is observed at the largest field that corresponds to the lowest spin-wave mode number $n = 1$. On the low-field side, the main resonance is accompanied by a series of peaks with a monotonously decreasing amplitude. Such a spin-wave spectrum is typical for confined circular nanodots \cite{Kak04apl}.
We observe that the two used approaches lead to shifts of the spin-wave resonance fields in opposite directions with respect to the reference state ($D=0$, $\tau=0$). At the same time, the shape and the intermodal distance pattern evolve consistently which is indicative of compositional uniformity and magnetic homogeneity of the samples. After integration and subtraction of the background, the experimental spectra were compared with multipeak Lorentzian functions to obtain the resonance fields for each mode. A theoretical model \cite{Kak04apl} was applied to fit the experimental data using $M_s$ and $A$ as two fitting parameters and assuming the gyromagnetic ratio of $\gamma/2\pi = 3.05$\,MHz/Oe \cite{Tok15prl}. The application of a least-square algorithm allowed us to deduce the magnetic parameters for all individual nanodisks with a precision of about $5$\%. Figure \ref{f3} illustrates that the best theoretical fits (solid lines) nicely describe the experimental data (symbols). We note that the location of the main resonance peak is primarily determined by $M_s$. The value of $A$ only weakly affects the position of the main resonance peak, however, it strongly affects the positions of the higher-order peaks.

\begin{figure}[t!]
    \centering
    \includegraphics[width=0.96\linewidth]{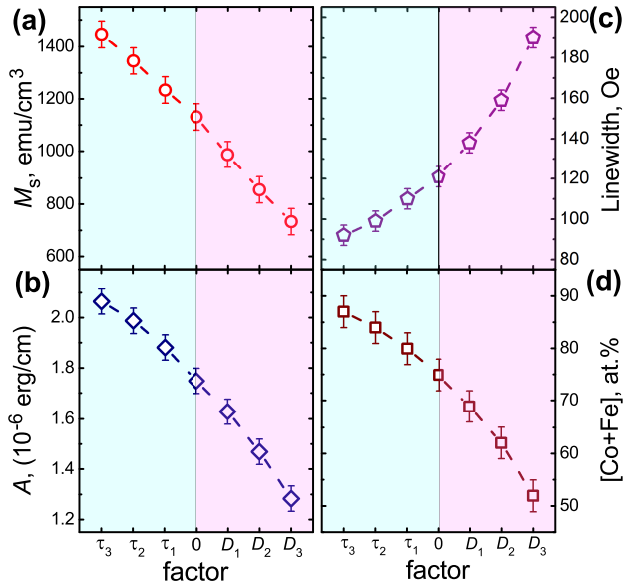}
    \caption{Evolution of the magnetization $M_s$ (a), the exchange constant $A$ (b), the linewidth (c) and the [Co+Fe] content with increase of the Ga ion irradiation dose (doses $D_1$ to $D_3$, light magenta background) and the electron beam waiting time ($\tau_1$ to $\tau_3$, light blue background). Dashed line are guides for the eye.}
    \label{f4}
\end{figure}

The deduced $M_s$ and $A$ values are reported in Fig. \ref{f4}(a,b). The field-sweep FMR linewidth, determined as the peak-to-peak distance in Fig. \ref{f2}(b), is presented in Fig. \ref{f4}(c). We next analyze their evolution in comparison with the composition of the disks inferred from energy-dispersive x-ray (EDX) spectroscopy. The EDX data in Fig. \ref{f4}(d) reveal an increase of the [Co+Fe] content from about $75\%$ at. in the as-grown sample ($\tau_0=0$) to about $87\%$ for the sample written with the beam parking time $\tau_3=50$\,ms, Fig. \ref{f4}(d). We attribute the increase of the metal content in the disk written with $\tau_3 = 50\,$ms to a nearly precursor depletion-free mode during its deposition, as the electron beam parking on such a timescale allows the precursor to replenish. The increase of the metal content correlates well with the increase of $M_s$ and $A$ and the decrease of the linewidth in Fig. \ref{f4}. In contrast, irradiation with Ga ions causes a degradation of the magnetic properties of the nanodisks, leading to a reduction of $M_s$ and $A$, and an increase of the linewidth.

The particular values of $\tau$ and $D$ were chosen as a scale factor in Fig. \ref{f4} to demonstrate in one plot the opposite character of the used approaches and the whole tuning range of $M_s$ and $A$ for Co-Fe nanostructures. The data in Fig. \ref{f4}(a) suggest that $M_s$ can be varied by a factor of about two, which offers sufficient flexibility, e.g. for the design of graded-index magnonic conduits \cite{Dav15prb,Whi19prb,Gru18prb} and magnonic crystals \cite{Kra14pcm,Chu17jpd,Zak20pcm}. The data in Fig. \ref{f4}(c,d) indicate that a decrease of the metal content in the disks by about $35$\,at.\% is accompanied by a factor-of-two linewidth broadening. Yet, we note that the linewidth ($90$\,Oe at 9.85\,GHz) in the most CoFe-rich disk is a factor of about two larger than in sputtered Py films \cite{Kal06jap}.

To summarize, we have demonstrated a methodology for the magnetization and exchange stiffness engineering in Co-Fe micrometer-sized nanodisks. The disks were fabricated by the direct-write nanofabrication technology of focused electron beam induced deposition. The analysis of the perpendicular FMR measurements data revealed an increase of the magnetization $M_s$ and the exchange stiffness $A$ in the disks written with longer e-beam waiting time and a reduction of $M_s$ and $A$ in disks irradiated with Ga ions. The physical reason for the larger $M_s$ and $A$ is the operation of FEBID in a nearly depletion-free precursor regime, which results in a higher metal content and smaller damping (which is proportional to the FMR linewidth) in the disks. The decrease of $M_s$ and $A$ in conjunction with the linewidth increase reflects a degradation of the magnetic properties and a higher inhomogeneity of the disks irradiated with Ga ions. Specifically, the achieved variation of $M_s$ from about $720$\,emu/cm$^3$ to about $1430$\,emu/cm$^3$ allows for its engineering in a broad range, continuously bridging the gap between the $M_s$ values of such widely used magnonic materials as Py and CoFeB \cite{Chu17jpd}. The $M_s$ tuning is accompanied by a variation of the exchange stiffness in the range $1.28\times10^{-6}$\,erg/cm to $2.07\times10^{-6}$\,erg/cm and the field-sweep FMR linewidth between $190$\,Oe and $90$\,Oe. The reported approach opens a way towards nanoscale 2D and 3D systems with controllable and space-varying magnetic properties.

\footnotesize{
OVD acknowledges the Austrian Science Fund (FWF) for support through Grant No. I 4889 (CurviMag).
The Portuguese team acknowledges the Network of Extreme Conditions Laboratories-NECL and Portuguese Foundation of Science and Technology (FCT) support through Project Nos. NORTE-01-0145-FEDER-022096, POCI-0145-FEDER-030085 (NOVAMAG), and EXPL/IF/00541/2015.
BB acknowledges financial support by the Vienna Doctoral School in Physics (VDSP).
KL and AVC acknowledge the Austrian Science Fund (FWF) for support through Grant No. I 4696.
KG acknowledges support by IKERBASQUE (the Basque Foundation for Science). The work of KG was supported by the Spanish Ministerio de Ciencia, Innovacion y Universidades grant FIS2016-78591-C3-3-R.
AVC and QW acknowledge support within the ERC Starting Grant no. 678309 MagnonCircuits.
Support through the Frankfurt Center of Electron Microscopy (FCEM) is gratefully acknowledged.
}
\vfill

\end{document}